\documentclass[]{cJAS2e}

\usepackage{subfigure}
\usepackage{hyperref}

\theoremstyle{plain}

\theoremstyle{remark}

\newcommand{\di}{\ensuremath{\,\mathrm{d}}}         

\newcommand{\arxiv}[1]{\href{http://arXiv.org/abs/#1}{arXiv: #1}}

\newcommand{\Poi}{\ensuremath{\text{Poi}}}          
\newcommand{\Bin}{\ensuremath{\text{Bin}}}          
\newcommand{\Mul}{\ensuremath{\text{Mul}}}          
\newcommand{\Dir}{\ensuremath{\text{Dir}}}          
\newcommand{\Gam}{\ensuremath{\text{Ga}}}           
\newcommand{\Be}{\ensuremath{\text{Be}}}            

\begin{document}

\title{Objective Bayesian analysis of counting experiments with
  correlated sources of background}

\author{Diego Casadei$^{a}$, Cornelius Grunwald$^{b}$, Kevin
  Kr{\"o}ninger$^{b}$$^{\ast}$\thanks{$^\ast$Corresponding
    author. Email: kevin.kroeninger@tu-dortmund.de
\vspace{6pt}}, and Florian Mentzel$^{b}$\\\vspace{6pt}
$^{a}${\em{Fachhochschule Nordwestschweiz, Bahnhofstrasse 6, 5210 Windisch, Switzerland, and University of Birmingham, UK}};
$^{b}${\em{Lehrstuhl Experimentelle Physik IV, TU Dortmund, Otto-Hahn-Stra{\ss}e~4a, 44227 Dortmund, Germany}} }

\maketitle

\begin{abstract}
  Searches for faint signals in counting experiments are often
  encountered in particle physics and astrophysics, as well as in
  other fields.  Many problems can be reduced to the case of a model
  with independent and Poisson-distributed signal and background.
  Often several background contributions are present at the same
  time, possibly correlated.  We provide the analytic solution of the
  statistical inference problem of estimating the signal in the
  presence of multiple backgrounds, in the framework of objective
  Bayes statistics.  The model can be written in the form of a
  product of a single Poisson distribution with a multinomial
  distribution.  The first is related to the total number of events,
  whereas the latter describes the fraction of events coming from
  each individual source.  Correlations among different backgrounds
  can be included in the inference problem by a suitable choice of
  the priors.
\end{abstract}

\begin{keywords}
  Counting experiments; Bayesian inference; objective priors
\end{keywords}


 \section{Introduction}

 Searches for faint signals in counting experiments are often
 encountered in particle physics (e.g.~in searches for new resonances)
 and astrophysics (e.g.~in searches for new sources of high-energy
 photons).  Many practical problems can be reduced to the case of a
 counting experiment where the events come from the sum of two
 independent Poisson-distributed contributions, usually referred to as
 ``signal'' and ``background''.  Although an accurate estimate of the
 background is desired in most applications, the sole parameter of
 interest is typically the signal intensity.  In other words, the
 background intensity can be viewed as a nuisance parameter of the
 statistical inference problem, whose goal it is to estimate the
 signal strength given the measured count rate.

 Sometimes the background is the result of different sources.  In such
 cases, one has a model in which the signal is superimposed to the
 total background.  If we knew which event comes from which source, we
 could keep separate lists for the number of observed events for each
 source of signal or background. However, in many practical problems,
 only fewer quantities are directly observable, and sometimes only the
 total number of observed events can be measured. An example is the
 number of events recorded by a Geiger-M{\"u}ller counter when
 measuring the intensity of a weak radioactive sample. The counts give
 only information about the total rate of radioactive decays, and not
 about the isotopic composition of the sample or the environment.
 Often one has prior knowledge about the individual contributions
 which can be used to improve the result of the inference. In the
 example of the weak radioactive source, this could be any knowledge
 about the rate of natural radioactivity of the surrounding and the
 rate of cosmic muons, obtained from a measurement without the source.

 In this paper we address the inference problem outlined above by
 means of objective Bayesian methods. These are based on the
 likelihood function, on informative background priors, and on an
 objective signal prior. This work is based on Refs.~\cite{PoiSigBkg}
 and \cite{PoiSigBkg2}, and extends their results.

 Choosing a Bayesian approach forces the scientist to be explicit
 about all the assumptions made.  Some of them may be ``subjective'',
 in the sense that they reflect the professional experience of the
 scientist.  As this is validated by numerous cross checks and by the
 criticism of the colleagues, a better term might be
 ``inter-subjective'' but we will keep the standard terminology.  On
 the other hand, ``objective'' means that the posterior depends only
 on the assumption of the model, although there is always some degree
 of (scientific) subjectivity in the choice of the model in the first
 place.  This is exactly the meaning attributed to ``objective'' in
 frequentist approaches.  Hence we are concerned here with Bayesian
 results that are by construction as objective (or subjective) as
 frequentist results.

 Bayesian solutions have the advantage of a very simple
 interpretation: one obtains a range of admissible values for the
 parameter of interest (the signal intensity) together with the
 probability that the true value belongs to that range.  On the other
 hand, the true value of the parameter is not a random quantity in the
 frequentist approach and thus one can not speak about the probability
 that it is inside some interval.  Hence, a frequentist solution is a
 statement about the probability of the data given the true value, in
 a picture in which the experiment is imagined to be identically
 repeated a large number of times. In practical situations, this is a
 very idealized picture, as identically repeating the experiment is
 either unfeasible or too expensive.  Furthermore, when feasible it
 might also appear undesirable.  For example, in high-energy physics
 data are collected during several ``runs'', ultimately related to the
 availability of particle interactions.  Colliders inject particles
 before the run starts and keep them circulating until the collision
 rate drops enough to justify a new fill.  Detectors enable data
 taking after each fill, and keep running until collisions are
 detected or some problem forces restarting the run.  In practice,
 although to some extent all runs could be considered as identical
 repetitions of the same experiment, during data analysis they are
 joined to form a bigger sample, rather then treating them as repeated
 experiments.

 From a computational perspective, Bayesian methods require the
 calculation of multidimensional integrals while frequentist methods
 typically adopt maximization procedures. The latter are typically
 less CPU-intensive, if those calculations are performed using
 numerical methods. However, we are concerned here with a problem that
 can be treated analytically, hence the computing time is negligible.

 Both approaches are perfectly legitimate.  When addressing
 statistical inference, it is important to clearly formulate the
 question one wants to answer, because this implies the choice of the
 paradigm.  For example, if one wants to speak in terms of the
 probability of a model (in our case, a model including contributions
 from a signal source) given the experimental result (e.g. the total
 count rate), then the Bayesian approach is the correct paradigm.  On
 the other hand, if the question is cast in terms of how (un)likely
 the present outcome would be in an ideal infinite sequence of
 identical repetitions of the same experiment, then the frequentist
 approach is the choice.
 We perform the statistical inference from the Bayesian point of view
 because we believe that this answers the most typical question asked
 by scientists once an experimental result is made public: what is
 the probability that this or that model is (in)correct?

 Regardless of using a Bayesian or frequentist approach, the
 statistical model used in the inference problem has to be defined
 carefully. Although the approach chosen in the following guarantees
 that in the asymptotic limit of a very large number of observed
 events the posterior will concentrate around the maximum of the
 likelihood, the most interesting class of problems is on the opposite
 side, i.e.~when collecting a very low (possibly null) number of
 events.  It is hence important that the general solution to this
 problem does not rely on asymptotic approximations and assumptions,
 e.g. by using a Gaussian model in the regime of small count rates.

 Considering the contributions of signal and different background
 sources to the count rate allows not only to infer on the
 signal-plus-background rate, but also to investigate the individual
 contributions. It is possible to model such a situation by either a
 product of independent Poisson distributions --- one for each signal
 or background source --- or by a product of a single Poisson
 distribution and a multinomial distribution. In the latter form we
 can encode correlations between the different contributions with the
 help of an informative Dirichlet prior. The inference problem may be
 then split into two parts: inference about the total intensity, and
 inference about the relative contributions of the different
 sources. The main result is that an analytic objective Bayesian
 solution can be obtained, which is valid for any number of measured
 events (even when it is very low or zero).

 An example for such a situation is the search for contributions of a
 new physics process at a hadron collider in a particular part of a
 spectrum, say, an invariant mass peak on top of a set of Standard
 Model background processes. The expected number of events due to
 background sources which are estimated based on theoretical
 cross-sections are partially correlated due to common uncertainties,
 for example the integrated luminosity of the data set or the scale of
 the invariant mass.  These correlations can be modeled by suitably
 choosing a prior for the multinomial part.

 This paper is organized as follows. Section~\S\ref{sec-likelihood}
 gives details about the statistical models, while
 Section~\S\ref{sec-bayes} describes the Bayesian
 inference. Section~\S\ref{sec-summary} concludes the paper.


 \section{The statistical model}\label{sec-likelihood}

 A statistical model is the specification of the probability to
 observe any outcome given the values of the parameters.  When
 considered as a function of the unknown parameters given the observed
 data, the model is called the likelihood function.

 We first address the case of a counting experiment with $k$
 contributions and for which the counts $X_1,X_2,\ldots,X_k \in
 \mathbb{N}$ are independent Poisson variables and are observed
 independently. We recall the well known result that the likelihood
 function can either be written as a product of Poisson distributions,
 or as the product of a Poisson distribution for the total number of
 counts, $N \equiv X_1+X_2+\ldots+X_k$, and a multinomial distribution
 describing how the counts are partitioned across the different
 contributions. 

 This is most often used when considering the entries in different
 histogram bins. The number of events in each bin, taken alone,
 follows a Poisson distribution. The shape of the histogram,
 normalized to unit area, is described by a multinomial distribution.
 For this reason, it is sometimes useful to imagine that our counts
 $X_1,\ldots,X_k$ are represented by a histogram with $k$ bins.

 Later, we consider the situation when one additional contribution has
 a special meaning, that is it represents the signal.

 \subsection{$k$ independent Poisson sources and $k$ observables}
 \label{sec-k-bins}

 If $X_i \in \mathbb{N}$ with $i=1,\ldots,k$ are independent Poisson
 random variables, the statistical model of the data is a product of
 Poisson distributions
 \begin{equation}
   \vec{X} \sim \prod_{i=1}^{k} \Poi(X_i|\lambda_i)
 \end{equation}
 with parameters $\lambda_i \in \mathbb{R}^+$, where
 \begin{equation}\label{eq-Poi}
   \Poi(X | \lambda) \equiv \frac{\lambda^{X}}{X!} \, e^{-\lambda}   \; .
 \end{equation}

 The outcome of the experiment is the $k$-tuple $\vec{x}
 \equiv(x_1,x_2,\ldots,x_k)$ of observed counts for each contribution,
 using a notation in which lowercase letters stand for actual values
 of the random variables denoted with uppercase letters.  The total
 number of observed events is $n \equiv \sum_i x_i$.

 One the other hand, the conditional distribution of the random vector
 $\vec{X} \equiv (X_1,X_2,\ldots,X_k)$, given the total number of
 counts $n \equiv \sum_{j=1}^{k} X_j$, is a multinomial distribution
 \begin{equation}\label{eq-Mul}
   \Mul(\vec{X} \, | n, \vec{\eta}) \equiv
          \frac{n!}{X_1! \, X_2! \, \cdots X_k!}
          \, \eta_1^{X_1} \, \eta_2^{X_2} \, \cdots \, \eta_k^{X_k}
 \end{equation}
 with the conditions
 \begin{equation}
   \sum_{j=1}^{k} X_j = n \, .
 \end{equation}
 and
 \begin{equation}\label{eq-Mul-cond}
   \sum_{j=1}^{k} \eta_j = 1 \, .
 \end{equation}
 The parameters $\vec{\eta} \equiv (\eta_1,\eta_2,\ldots,\eta_k)$
 represent the relative proportions of the Poisson rates:
 \begin{equation}\label{eq-proportions}
   \eta_i \equiv \frac{\lambda_i}{\lambda}
   \quad \text{with} \quad
   \lambda \equiv \sum_{j=1}^{k} \lambda_j    \; .
 \end{equation}
%
 The condition~\eqref{eq-Mul-cond} implies that there are only $k-1$
 independent shape variables in Eqn.~\eqref{eq-Mul}, and could be used
 to write the multinomial distribution in a less symmetric form, for
 example by replacing $\eta_k$ with $1-\sum_{j=1}^{k-1} \eta_j$ and
 $X_k$ with $n-\sum_{j=1}^{k-1} X_j$ in Eqn.~\eqref{eq-Mul}.

 The binomial distribution is a special case of the multinomial
 distribution with $k=2$,
 \begin{equation}
   \label{eq-bin}
   \begin{split}
     \Bin(Y|\,n,\varepsilon) &\equiv \frac{n!}{Y! \, (n-Y)!} \,
                            \varepsilon^{Y} (1-\varepsilon)^{n-Y}
    \\
      &= \frac{n!}{X_1! \, X_2!}
          \, \eta_1^{X_1} \, \eta_2^{X_2}
       = \Mul(\vec{X} \, | n, \vec{\eta})
     \end{split}
 \end{equation}
 with
 \begin{equation}
    \vec{X} \equiv (X_1,X_2) = (Y, n-Y) \;, \quad
    \vec{\eta} \equiv (\eta_1, \eta_2) = (\varepsilon, 1-\varepsilon)
 \end{equation}

 A well-known result is that the unconditional distribution of
 $\vec{X}$ can be factored into the product
 \begin{equation}\label{eq-model-k-bins}
   \vec{X} \sim \prod_{i=1}^{k} \Poi(X_i|\lambda_i)
           = \Poi(n | \lambda) \;
             \Mul(\vec{X} \, | n, \vec{\eta})
 \end{equation}
 of a single Poisson term and a multinomial distribution.  The Poisson
 distribution gives the probability of observing a total number of
 events $n$, given the sum of all expected contributions, $\lambda$.
 The multinomial term describes how the individual counts $X_i$
 distribute among the $k$ bins.

 Let us now assume that we have measured a total of $n$ counts, and
 that we want to infer about the model parameter using
 Eqn.~\eqref{eq-model-k-bins} as the likelihood function.  An immediate
 consequence of Eqn.~\eqref{eq-model-k-bins} is that $n$ carries no
 information about $\vec{\eta}$ (and vice versa), but only about
 $\lambda=\sum_{j=1}^{k} \lambda_j$. It is no surprise that the
 statistical inference about the total rate only requires the
 total number of counts.

 On the other hand, $\vec{\eta}$ encodes the shape of the distribution
 of counts across the $k$ contributions, but not the normalization:
 likelihood-based inferences about $\vec{\eta}$ are the same whether
 one regards $(x_1,x_2,\ldots,x_k)$ as sampled from $k$ independent
 Poisson sources or from a single multinomial process.  In other
 words, estimates and tests of any function of $\vec{\eta}$ will be
 the same whether we regard the number of counts as a random variable
 or as a fixed value.

 This means that we can separately address the statistical inference
 problems related to the total count rate and that related to the
 contributions from different background sources. The first step is to
 solve a problem with a single Poisson model and $n$ total observed
 events.  The second step is to solve a multinomial problem which
 focuses on the distribution of counts, given the total number $n$ of
 observed counts.


 \subsection{Signal in the presence of $k$ background sources}
 \label{sec-sig-k-bkg}

 If a Poisson distributed signal source exists on top of $k$
 background sources, and is independent of the background, one of the
 $k+1$ contributions to the observed number of counts assumes a
 special meaning (the signal) and the model can be written as
 \begin{equation}
   \label{eq-sig-and-K-bkg}
   \Poi(Y|s) \, \prod_{i=1}^{k} \Poi(X_i|b_i) =
     \Poi(n | s+b) \; \Mul(\vec{Z} \, | n, \vec{\zeta}) = P(n,\vec{Z}|s,\vec{b})
 \end{equation}
 where 
 \begin{eqnarray*}
   b \equiv \sum_{i=1}^k b_i \;,\quad
   & \displaystyle
   X \equiv \sum_{i=1}^k X_i \;,\quad
   &
   n \equiv Y+X  \;,
   \\
   \vec{Z} \equiv (Y, \vec{X}) \;,\quad
   &
   \quad
   &
   \vec{\zeta} \equiv \left(\frac{s}{s+b},
                    \frac{b_1}{s+b}, \ldots, \frac{b_k}{s+b}\right)  
\, .
 \end{eqnarray*}

Making use of the interpretation of a histogram described earlier, the
different counts $\vec{Z}$ can be viewed as a histogram with counts
from one signal region and $k$ auxiliary measurements.

\subsection{The multi-channel signal + background problem}

 We are now interested in the case that the data show an evidence
 that, in addition to the known ``background'' source with
 non-negative bin yields $\vec{b} \equiv (b_1,b_2,\ldots,b_k)$ and
 overall strength $b \equiv \sum_{j=1}^{k}b_j$ representing nuisance
 (i.e.~not interesting) parameters, there is a non-null contribution
 from a ``signal'' source whose non-negative bin yields $\vec{s}
 \equiv (s_1,s_2,\ldots,s_k)$ and overall strength $s \equiv
 \sum_{j=1}^{k}s_j$ are our parameters of interest.  This is not just
 the ``discovery'' problem, for which one would typically focus on the
 total number of observed counts, because here we are also interested
 in the shape of the signal contribution.  This is likely to be the
 second step after a discovery phase of a new particle, for example
 when one wants to conduct a measurement of the mass or width of the
 new particle, or to distinguish between two competing models which
 predict different shapes for the distribution of some kinematical
 quantity.

 Even though we have $2k$ independent sources, we do not know if any
 given event has come from a signal or background source.  Hence we
 only have $k$ observables, the counts in $k$ bins.  Formally, we
 start by considering the signal random variables $Y_1,Y_2,\ldots,Y_k
 \in \mathbb{N}$, which are Poisson distributed $Y_i \sim \Poi(s_i)$,
 and the background variables $Z_1,Z_2,\ldots,Z_k \in \mathbb{N}$,
 which are also Poisson distributed $Z_i \sim \Poi(b_i)$.  Next, we
 define the random variables $X_i \equiv Y_i + Z_i$ which correspond
 to the observable counts in the $k$ bins.  As the sum of Poisson
 variables is again a Poisson variable, we have $X_i \sim
 \Poi(s_i+b_i)$, i.e.~we recover the notation used in
 section~\ref{sec-k-bins} above with $\lambda_i = s_i + b_i$, $\forall
 i=1,\ldots,k$ and $\lambda = s + b$.

 In terms of the variables $\vec{Y}$ and $\vec{Z}$, if $n$ total events
 have been observed the likelihood function for $2k$ bins is
\begin{equation}
   \Mul(\vec{Y}, \vec{Z} \,|\, n, \, s_1/s, \ldots, s_k/s,
                                  \, b_1/b, \ldots, b_k/b)
\end{equation}
 Now we partition the counts into $Y = \sum_{i=1}^k Y_i$ signal
 events and $n-Y = \sum_{i=1}^k Z_i$ background events (see Appendix). 
 Conditional on $Y$ (and on $n$), we can now write the previous
 multinomial as the product
\begin{equation}
   \Mul(\vec{Y} \,|\, Y, s_1/s, \ldots, s_k/s) \;
   \Mul(\vec{Z} \,|\, n-Y, b_1/b, \ldots, b_k/b)
\end{equation}
 Because we actually do not know $Y$, we need to sum over all possible
 ways of obtaining $Y$ signal events out of the total $n$ counts, with
 the help of a binomial distribution
 \begin{equation}\label{eq-Bin}
   \Bin(Y|\,n,\varepsilon) \equiv \frac{n!}{Y! \, (n-Y)!} \,
                               \varepsilon^{Y} (1-\varepsilon)^{n-Y}
 \end{equation}
 whose parameter $\varepsilon=s/(s+b)$ is the probability of obtaining a
 signal event when both signal and background are active sources.

 The likelihood for $2k$ bins is then
\begin{equation}\label{eq-2k-likelihood}
  \begin{split}
    L = \, \Poi(n \,|\, s+b) \; \sum_{Y=0}^{n}
     & \Bin(Y|\,n,s/(s+b)) \; \times
\\
     &\hspace{-2ex} \times \Mul(\vec{Y} \,|\, Y, s_1/s, \ldots, s_k/s) \; \times
\\
     &\hspace{-2ex} \times \Mul(\vec{Z} \,|\, n-Y, b_1/b, \ldots, b_k/b)
  \end{split}
\end{equation}
 As a cross check, we can count the degrees of freedom: 1 Poisson
 variable, plus 1 binomial variable, plus $k-1$ multinomial variables
 for the signal, plus $k-1$ multinomial variables for the background.
 In total, we have $2k$ variables: $s$ together with $k-1$ signal yields
 (e.g.~$s_1,\ldots,s_{k-1}$) are the parameters of interest, whereas
 the $k$ background yields $b_1,\ldots,b_k$ are nuisance parameters.

 We would have obtained the same result,
 equation~(\ref{eq-2k-likelihood}), by considering the
 likelihood~(\ref{eq-model-k-bins}) as the merged version of separate signal and
 background sources (see Appendix), and defining $\lambda_i=s_i+b_i$, such that
 \[
   \eta_i = \frac{s_i+b_i}{(\sum_{j=1}^{k} s_j) + (\sum_{j=1}^{k} b_j)}
           = \frac{s_i}{s+b}
           + \frac{b_i}{s+b}
 \]
 (one must not forget the binomial weights due to the splitting of the
 $n$ events into two unknown partitions).

 Finally, because we can only observe the sum of signal and background
 counts in each bin, we have to merge the latter into a set of $k$
 bins.  Given the observed $X_i$ counts in bin $i$, there are several
 ways in which signal and background events can give this sum.  Again,
 we account for this with the help of binomial weights,
 $\Bin(Y_i|\,X_i,\varepsilon_i)$ with $\varepsilon_i = s_i/(s_i+b_i)$.  The full
 likelihood becomes
\begin{equation}\label{eq-full-likelihood}
  \begin{split}
    L = \, & \Poi(n \,|\, s+b) \;
        \Mul\left(\vec{X} \,\Big|\, n,
           \frac{s_1+b_1}{s+b}, \ldots, \frac{s_k+b_k}{s+b}\right) \; \times
\\
           & \times  \left\{ \sum_{Y=0}^{n}
           \Bin\left(Y\Big|\,n,\frac{s}{s+b}\right) \right.
         \; \times
\\
           &\hspace{3ex} \times  \sum_{\vec{Y}_s=0 \,|\,Y}^{\vec{X}} \left[
           \Mul\left(\vec{Y} \,\Big|\, Y, 
             \frac{s_1}{s}, \ldots, \frac{s_k}{s}\right) \;
           \left. \prod_{i=1}^{k}
           \Bin\left(Y_i\Big|\,X_i,\frac{s_i}{s_i+b_i}\right)
           \right] \right\}
  \end{split}
\end{equation}
 that is the product of (\ref{eq-model-k-bins}) --- which is the first line in
 (\ref{eq-full-likelihood}) --- with terms accounting for all 
 possible combinations of $k$-tuples of signal counts and $k$-tuples
 of background counts which give as a result the observed counts
 $(X_1,\ldots,X_k)$ in each bin.

 In general, given the experimental result $(X_1,\ldots,X_k)$, the
 likelihood function~(\ref{eq-full-likelihood}) is the starting point
 for performing statistical inference about the parameters of
 interest, which are the signal yields $s_1,\ldots,s_k$.  However,
 when $k$ is not small this model is computationally very intensive,
 as the number of possible combinations becomes huge. 


 \section{Bayesian statistical inference}\label{sec-bayes}

 In the Bayesian approach, the statistical inference relies on the use
 of Bayes' theorem, which can be interpreted as the algorithm for
 updating our knowledge about a model and its parameters (modeled as a
 probability distribution) in view of the outcome of an experiment.
 The prior knowledge is encoded into the prior distribution for the
 parameters of the model, and their joint posterior distribution
 (obtained with Bayes' theorem) reflects our new state of knowledge
 about them.  In many cases, we are only interested in a subset of the
 parameters. One can obtain their joint marginal posterior probability
 density by integrating the full posterior over all other (nuisance)
 parameters.

 \subsection{Inference about $k$ independent Poisson sources and $k$ observables}

 We first address the problem of $k$ independent Poisson sources and
 $k$ observable count rates. The likelihood function is given by
 Eqn.~\eqref{eq-model-k-bins} with lowercase letters that remind us
 that the likelihood is a function of the parameters with fixed (and
 known) experimental outcome.

 Bayes' theorem gives the joint posterior probability density for
 $\lambda$ and $\vec{\eta}$
 \begin{equation}
   \label{eq-ksourcekobs}
   \begin{split}
   p(\lambda,\vec{\eta} \,|\, n,\vec{x}) & \propto
      p(\lambda \,|\, n)  \; p(\vec{\eta} \,|n, \vec{x})
\\ 
      & \propto [\Poi(n | \lambda) \; p(\lambda)] \;
      [\Mul(\vec{x} \, | n, \vec{\eta}) \; p(\vec{\eta})]
   \end{split}
 \end{equation}
 (where the normalization constant can be found by integrating over
 the r.h.s.)~in the form of a product of two posterior densities.  The
 first corresponds to a Poisson model with expectation value $\lambda$
 and prior density $p(\lambda)$, from which $n$ events are generated.
 The second corresponds to a multinomial model with parameters
 $\vec{\eta}$ and prior density $p(\vec{\eta})$, generating the
 observed vector $\vec{x}$ of counts.  They are two inference problems
 which can be solved independently, as is shown below.

 \subsubsection{Prior knowledge about the Poisson problem}

 \subparagraph{Conjugate prior} 

 The most convenient functional form for the prior density of the
 Poisson parameter is a Gamma function \cite{PoiSigBkg}, i.e., the
 conjugate prior for a Poisson model, $\Poi(n | \lambda)$,
\begin{equation}\label{eq-gamma}
   \Gam(\lambda\,|\,\alpha,\beta)
              \equiv \frac{\beta^\alpha}{\Gamma(\alpha)}
                \, \lambda^{\alpha-1} \, e^{-\beta \lambda}
\end{equation}
 with shape parameter $\alpha>0$ and rate parameter $\beta>0$ (or
 scale parameter $\theta=1/\beta>0$).
 When the prior is $p(\lambda) = \Gam(\lambda\,|\,\alpha,\beta)$, the
 posterior is
 \begin{equation}\label{eq-Poi-post}
   p(\lambda \,|\, n) = \Gam(\lambda\,|\,\alpha+n,\beta+1)
   \; .
\end{equation}

 \subparagraph{Informative prior}
 
In the simple but frequent case in which the prior knowledge about
 $\lambda$ is summarized by its expectation $E[\lambda]$ and variance
 $V[\lambda]$, the Gamma parameters are determined with the method of
 moments by imposing $E[\lambda] = \alpha/\beta$ and $V[\lambda] =
 \alpha/\beta^2$.  This gives $\beta = E[\lambda]/V[\lambda]$ and
 $\alpha = \beta\, E[\lambda]$.  Alternatively, one could start from
 the prior most probable value (the Gamma mode is at
 $(\alpha-1)/\beta$ for $\alpha > 1$) and variance, or from the
 knowledge of intervals covering given prior probabilities
 (e.g.~68.3\% or 95\% probability; this requires a numerical treatment
 to find $\alpha$ and $\beta$), or from any set of conditions which is
 sufficient to determine the shape and rate parameters.

 In principle, if there is quite detailed information about the prior
 for $\lambda$ \emph{and} it is known that the Gamma density does not
 correctly represent its shape, one has two options.  The first one is
 to adopt a different functional form and solve Bayes' theorem
 numerically, and the second is to find a linear combination of Gamma
 densities which approximates the prior well enough.  In the latter
 case, the posterior will be a linear combination of Gamma functions.
 This simplifies the treatment without biasing the result, provided
 that enough terms are present in the linear combination of priors,
 and it is hence preferable to numerical integration.  When $n$ is
 large enough, the solution becomes lesser and lesser dependent on the
 exact shape of the prior,\footnote{Bayesian statistics obey the
   maximum likelihood ``principle'' (which is indeed a theorem in this
   framework): when $n$ is arbitrarily large the shape of the prior
   does not matter at all, provided that it is strictly non-null in
   the region where the true value is.} hence one should consider
 using numerical methods or a linear combinations only if $n$ is very
 small and --- most importantly --- when one is very confident that a single
 Gamma density is not good enough to model the prior.  In practical
 problems, this happens very rarely (although it is not impossible),
 and a vague knowledge of the prior shape is typically sufficient.

 In cases where only vague prior information is available, it is
 recommended to assess the dependence of the posterior on the choice
 of the prior by comparing the results obtained with different priors,
 all compatible with the available information (e.g.~different
 functional forms giving the same first two moments).  When the
 posterior does not change appreciably, it means that $n$ is large
 enough for the likelihood to dominate and that the result is robust.
 Otherwise, one has to conclude that the experiment did not yield
 enough events for obtaining a solution that does not depend on our
 prior state of knowledge.  In this case, it is best to report the
 differences of a few key figures of merit (like the posterior mean
 and variance, or posterior credible intervals with given probability)
 which correspond to the choice of quite different priors.

 How different should the priors then be?  Of course, comparing our
 informative Gamma density with a delta-function does not make any
 sense: there are requirements that admissible priors must satisfy to
 be considered acceptable.  Luckily enough, such requirements are
 really minimal and typically reasonable, like requiring that the
 prior is strictly positive over the entire domain and that the
 posterior be a proper density (i.e.~it integrates to one).  In
 addition, a formal procedure exists which gives us the ``most
 distant'' prior to our informative prior.  We can call it the ``least
 informative'' prior or the ``objective'' prior, and it is provided by
 the reference analysis \cite{bernardo05} based on information theory.
 This reference prior is the function which maximizes the amount of
 missing information about the system, hence it is the ``most
 distant'' choice from our informative prior.  In other words, the
 reference prior encodes the minimal set of information about the
 problem.  Indeed, it only depends on the description of the model
 itself (in the case under consideration here, on the Poisson
 distribution) and makes no use of any prior knowledge about the
 parameter other than the specification of its domain (which for a
 Poisson problem is the entire positive real line).  For this reason,
 the reference prior shall be used when one wishes to adopt an
 ``objective prior''.

\subparagraph{Objective prior}

 When assessing the dependence of the result from the prior
 information, it is recommended to compare against the reference
 posterior \cite{bernardo05}, i.e.~the solution obtained with the
 reference prior $\pi(\lambda)$, which for the Poisson model coincides
 with Jeffreys' prior: $\pi(\lambda) = \lambda^{-1/2}$.  The reference
 posterior is then 
 \begin{equation}\label{eq-Poi-ref-post}
    p(\lambda \,|\, n) = \Gam(\lambda\,|\,n+\tfrac{1}{2},1)
 \end{equation}
 and also represents the best choice when one is required to report an
 ``objective'' result, or when one claims to have the minimal prior
 information about the Poisson parameter.  Compared to the informative
 prior, which is chosen to best encode all available prior
 information, the reference prior is the ``most distant'' choice from
 the information-theory point of view, as it encodes the minimal prior
 information.

 \subparagraph{A simple example}

 In an attempt to measure the amount of background activity in a new
 lab, one performs two subsequent measurements for the same
 duration. For the first measurement, no prior knowledge about the
 expected rate $\lambda$ is assumed, hence the non-informative prior is
 chosen. The measurement yields an observed number of events of
 $n_{1}=9$, so that the posterior distribution is $p(\lambda \,|\, 9)
 = \Gam(\lambda\,|\,9.5,1)$ with expectation value $E[\lambda]=9.5$
 and variance $V[\lambda]=9.5$.

 The posterior of the first measurement is then used as an informative
 prior for the second measurement.  The number of observed events is
 $n_{2}=12$ and so the posterior distribution is $p(\lambda \,|\, 12)
 = \Gam(\lambda\,|\,21.5,2)$ with expectation value $E[\lambda]=10.75$
 and variance $V[\lambda]=5.38$.

 One obtains the same result by merging the two data sets and applying
 the objective prior to this new ``experiment''.  In this case, the
 posterior is $p(\lambda' \,|\, 9+12) = \Gam(\lambda'\,|\,21.5,1)$,
 where the new random variable $\lambda' = 2 \lambda$ because merging
 the two observations is equivalent to observing for a duration twice
 as long.  It is very simple to show that, after changing variable,
 one obtains $\Gam(\lambda'\,|\,a,b) \di\lambda' =
 \Gam(\lambda\,|\,a,2b) \di\lambda$, which means that the posterior
 for the original parameter is $\Gam(\lambda\,|\,21.5,2)$, the same as
 above.

 \subsubsection{Prior knowledge about the multinomial problem}\label{sec-mul-post}

 \subparagraph{Conjugate prior}

 The conjugate prior of the multinomial model, $\Mul(\vec{x} \, | n,
 \vec{\eta})$, is the Dirichlet distribution, which means that the
 posterior is also a Dirichlet distribution and its parameters are
 simple functions of the prior and multinomial parameters.

 The Dirichlet distribution with concentration parameters $\vec{a}
 \equiv (a_1,\ldots,a_k)$ (with $k\ge2$ and $a_i>0$ for $i=1,\ldots,k$) is
 \begin{equation}\label{eq-Dir}
      \Dir(\vec{\eta} \,|\, \vec{a}) \equiv \frac{1}{B(\vec{a})} \,
                              \prod_{i=1}^{k} \eta_{i}^{a_i-1}
   \;,
 \end{equation}
 where the multidimensional Beta function has been used to write the
 normalization constant
 \begin{equation}\label{eq-Beta-nD}
   B(\vec{a}) \equiv \frac{\prod_{i=1}^{k} \Gamma(a_i)}{\Gamma(A)}
   \quad \text{with} \quad
   A \equiv \sum_{i=1}^{k} a_i
   \;.
 \end{equation}
 If the prior for the multinomial problem is $p(\vec{\eta}) =
 \Dir(\vec{\eta} \,|\, \vec{a})$, the posterior is a Dirichlet
 distribution with parameters $\vec{x}+\vec{a}$:
 \begin{equation}\label{eq-Dir-post}
   p(\vec{\eta} \, |\, n, \vec{x}) = 
      \Dir(\vec{\eta} \,|\, \vec{x}+\vec{a})
          = \frac{\prod_{i=1}^{k} \eta_{i}^{x_i+a_i-1}}{B(\vec{x}+\vec{a})}
 \end{equation}

 When $k=2$, the multinomial becomes a binomial distribution.  The
 corresponding conjugate prior is the Beta density,
 \begin{equation}
   \label{eq-Beta-dist}
   \Be(\varepsilon \,|\,a_1,a_2) \equiv \frac{1}{B(a_1,a_2)}
                        \varepsilon^{a_1-1} (1-\varepsilon)^{a_2-1}
 \end{equation}
 which is indeed the special case of a Dirichlet distribution with
 $k=2$, $\eta_1=\varepsilon$, $\eta_2=1-\varepsilon$.

 One obtains a Beta density also as the marginal distribution of any
 Dirichlet parameter, i.e.~the Beta density is the result of
 integrating a Dirichlet over $k-2$ parameters.

 \subparagraph{Informative prior}

 The following properties of the Dirichlet distribution can be used to
 find the values of the concentration parameters which best represent
 all the available prior information:
 \begin{eqnarray}
   E[\eta_i] &=& a_i / A
   \\
   \text{mode}[\eta_i] &=& \frac{a_i - 1}{A-k}
   \\
   V[\eta_i] &=& \frac{a_i (A - a_i)}{A^2 (A+1)}
   \\
   \text{Cov}[\eta_i,\eta_j] &=& \frac{-a_i a_j}{A^2 (A+1)}
 \end{eqnarray}

 Often the parameters may be grouped into two classes: a set of
 parameters that are independent from any other parameter, and a set
 of parameters that have pairwise correlations.  By grouping them
 correspondingly, one obtains a block-diagonal matrix in which the
 first block is actually diagonal.  Parameters in the first block are
 then fixed by imposing their prior variance and expectation.  For the
 others, one has a system of requirements, for example in the form of
 a collection of variances and covariances.  As there may be more
 constraints than free variables, the prior concentration parameters
 may be found e.g.~by solving a minimization problem.
 Alternatively, one may choose to solve the problem multiple times,
 each with a different choice of concentration parameters which
 satisfies $k$ independent constraints.

 On the other hand, if there are less constraints than the number of
 parameters\footnote{We expect this case to be just an academic
   exercise, as in all important problems that we are aware of all
   background components are known to some extent.}, one may impose
 the ``objective'' choice on the unconstrained variables, i.e.~equal
 prior concentration parameters.  This might be done in a two-step
 algorithm starting with an objective prior with concentration
 parameters all set to $0.8/k$ (see below), and by imagining an intermediate
 experiment in which only a subset of $m<k$ variables get updated to
 $\Dir(y_1+0.8/k,\ldots,y_m+0.8/k,0.8/k,\ldots,0.8/k)$.  The next step
 is to find real values $y_1,\ldots,y_m$ such that the first $m$
 concentration parameters satisfy all available constraints.  The
 result can be used as the prior for the actual experiment.

 \subparagraph{Non-informative prior}

 The problem of finding an objective prior for all multinomial
 parameters was addressed by \citet{BergerBernardo1992} and
 \citet{BBS2013}.  Being aware that for multidimensional models the
 reference prior depends on the ordering of the parameters,
 \citet{BBS2013} argued that, when no priority ranking is desired or
 allowed (in other words, when one wants all $\eta_i$ parameters to be
 on the same footing), some sort of ``overall objective prior'' shall
 be used which treats all parameters fairly well.  A very reasonable
 requirement is that the marginal posterior for any given parameter
 should be as close as possible to the solution obtained with the
 marginal model and the reference prior for that single parameter.

 For the multinomial model, the reference posterior when only $\eta_i$
 is of interest is $\Be(\eta_i \,|\, x_i + 1/2, n - x_i +
 1/2)$.\footnote{As the multinomial with $k=2$ is equivalent to a
   binomial distribution, this result can also be proved with the
   latter model.  The reference posterior is the same Beta density, as
   it is shown for example in Ref.~\cite{efficiency}.}  On the other
 hand, if the prior $\Dir(\vec{\eta}\,|\,a,a,\ldots,a)$ is used ---
 using the same concentration parameter $a$ for all multinomial
 $\eta_i$ parameters, as they are treated on the same footing ---, the
 marginal posterior for $\eta_i$ is instead $\Be(\eta_i \,|\, x_i + a,
 n - x_i + (k-1)a)$.  The goal is to find $a$ such that, in an average
 sense, this result is closest to $\Be(\eta_i \,|\, x_i + 1/2, n - x_i
 + 1/2)$ for each $i$.  Among different possible approaches to this
 problem \cite{BBS2012}, \citet{BBS2013} addressed it with a
 hierarchical approach and showed that the overall objective prior is
 a Dirichlet density with all identical concentration parameters,
 whose value is very well approximated by the simple choice $a^* =
 0.8/k$ (just a bit lower than the intuitive choice $1/k$), which
 gives an objective Dirichlet posterior with parameters $x_i+0.8/k$.

 \subparagraph{A simple example, continued}
 
 In the case of two subsequent measurements, one might be interested
 in the background stability with time.  This is relevant during the
 preparation of low-background experiments, e.g.~if newly produced
 parts of an experiment are brought into underground laboratories.  In
 our example, we could imagine to have installed new equipment between
 the two measurements and ask ourselves if the somewhat larger number
 of counts in the second measurement suggests that the equipment has
 increased the overall background rate.  One way of checking this is
 to look at the 2-bins histogram with $x_1 = 9$ and $x_2 = 12$ counts
 and quantify the departure from equipartition.  A more formal
 procedure would be to test against the null hypothesis that the
 background rate is constant, but it is not necessary in this
 particular example, which is quite straightforward.

 To better illustrate the problem, we use first a binomial model with
 $y=x_1=9$ successes out of $n=x_1+x_2=21$ total trials and look at
 the deviation from $\varepsilon=0.5$, where $\varepsilon$ is the
 ratio between the initial and the total rate.  The reference
 posterior for $\varepsilon$ is \cite{efficiency}
 \begin{equation} 
   \label{ex1-ref-post1}
   p(\varepsilon|y,n) = \Be(\varepsilon|9.5,12.5)
 \end{equation} 
 The posterior mean, mode and standard deviation are
 \begin{equation} 
   \label{ex1-ref-post1-summary}
   E[\varepsilon] = 0.4318  \qquad
   M[\varepsilon] = 0.4250  \qquad
   \sigma[\varepsilon] = 0.1033
 \end{equation} 

 If we start with $\Mul(x_1,x_2|\varepsilon,1-\varepsilon)$ and use
 the overall objective prior $\Dir(\varepsilon,1-\varepsilon|0.4,0.4)$
 recommended by \citet{BBS2013}, we obtain instead the marginal posterior
 \begin{equation} 
   \label{ex1-ref-post2}
   p(\varepsilon|y,n) = \Be(\varepsilon|9.4,12.4)
 \end{equation} 
 whose mean, mode and standard deviation are
 \begin{equation} 
   \label{ex1-ref-post2-summary}
   E[\varepsilon] = 0.4312  \qquad
   M[\varepsilon] = 0.4242  \qquad
   \sigma[\varepsilon] = 0.1037
 \end{equation} 
 Thus the two alternative solutions are hardly distinguishable in practice.

 The peaks are $0.726\sigma$ and $0.730\sigma$ lower than
 $\varepsilon=0.5$, not a significant difference to conclude that the
 background level has changed between the two measurements, even
 without performing a formal test.  Hence it is safe to use both
 measurements to estimate the background level in the cavern.

 \subparagraph{An example from high-energy physics: lepton universality}

 Assume an experiment built to test lepton universality (see,
 e.g. Refs.~\cite{Agashe:2014kda} for the current experimental status)
 in $Z$-boson production which is equally sensitive to electrons,
 muons and taus, i.e. assuming equal reconstruction efficiencies,
 momentum resolutions and so on. Events containing $Z$-bosons are
 selected by measuring final states with two opposite-charge leptons
 and by requiring their invariant mass to be consistent with the mass
 of a $Z$-boson. Adding further stringent event-selection
 requirements, the resulting event sample can be made practically
 background-free. The number of observed events in this example are
 $n_{e}=17$, $n_{\mu}=19$ and $n_{\tau}=12$. Lepton universality is
 tested by estimating the branching ratios, e.g., $\eta_{e}=n_{e}/n$,
 where $n=n_{e}+n_{\mu}+n_{\tau}=48$, and comparing them. If they are
 found to not be equal, then lepton universality is disfavored.

 Using Eqn.~\eqref{eq-ksourcekobs} with a reference prior for the
 overall rate $\lambda$ and an objective prior for the multinomial
 part yields
 \begin{eqnarray*}
   p(\lambda,\vec{\eta} \,|\, n,\vec{x}) & \propto & 
   [\Poi(n | \lambda) \; \lambda^{-1/2}] \; [\Mul(\vec{x} \, | n, \eta_{e}, \eta_{\mu},\eta_{\tau}) \; \Dir(\eta_{e}, \eta_{\mu},\eta_{\tau} \,|\, 0.8/3, 0.8/3, 0.8/3)] \, .
 \end{eqnarray*}

\begin{figure}[t]
   \begin{center}
     \begin{tabular}{ccc}
       \includegraphics[width=0.30\textwidth]{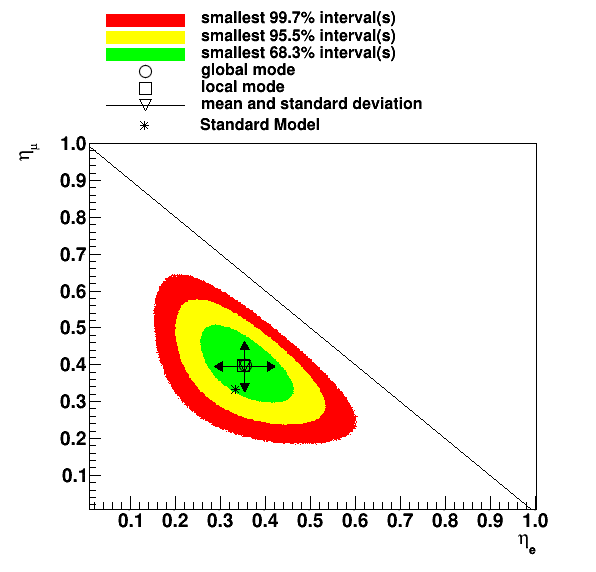} & 
       \includegraphics[width=0.30\textwidth]{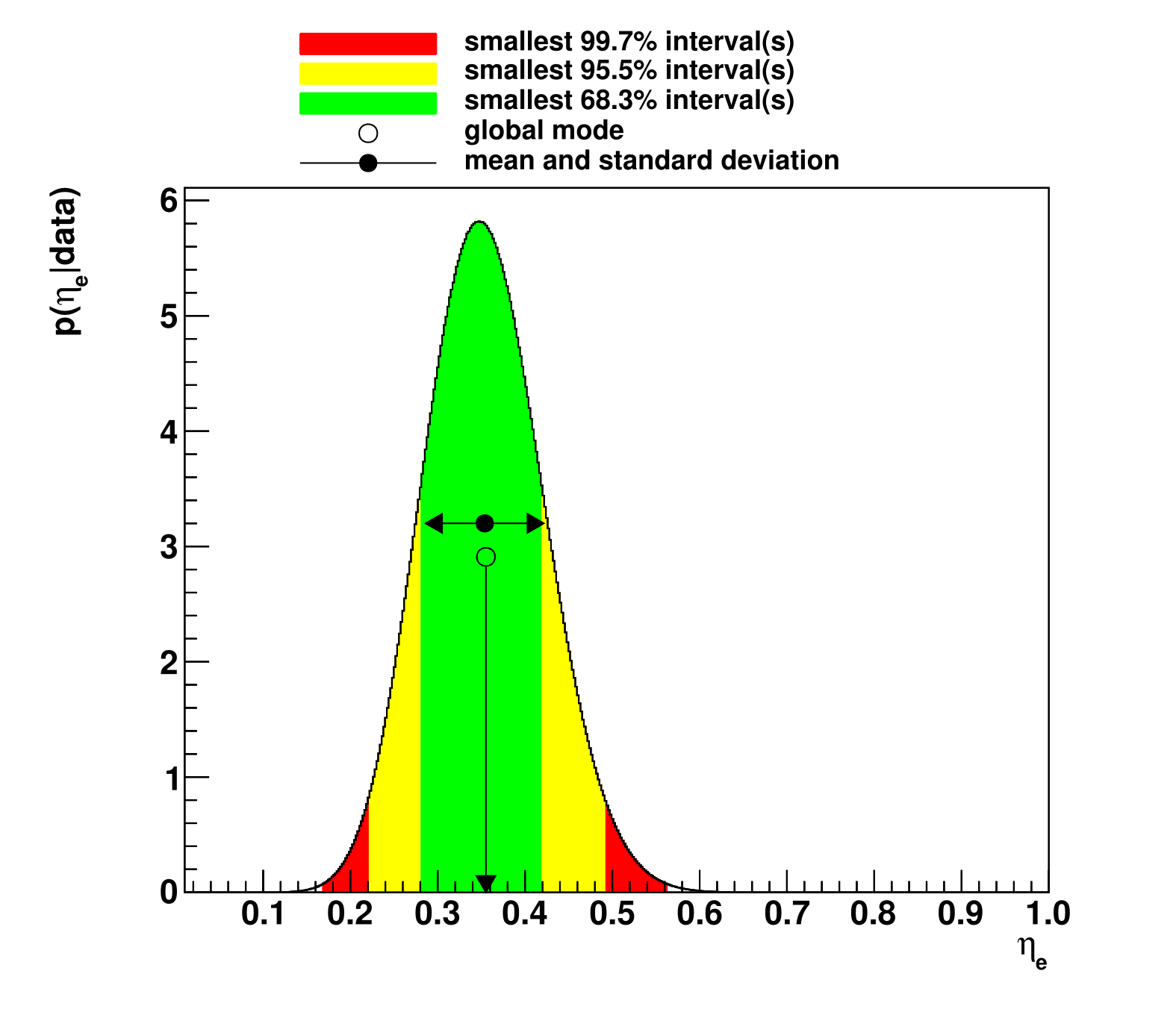} & 
       \includegraphics[width=0.30\textwidth]{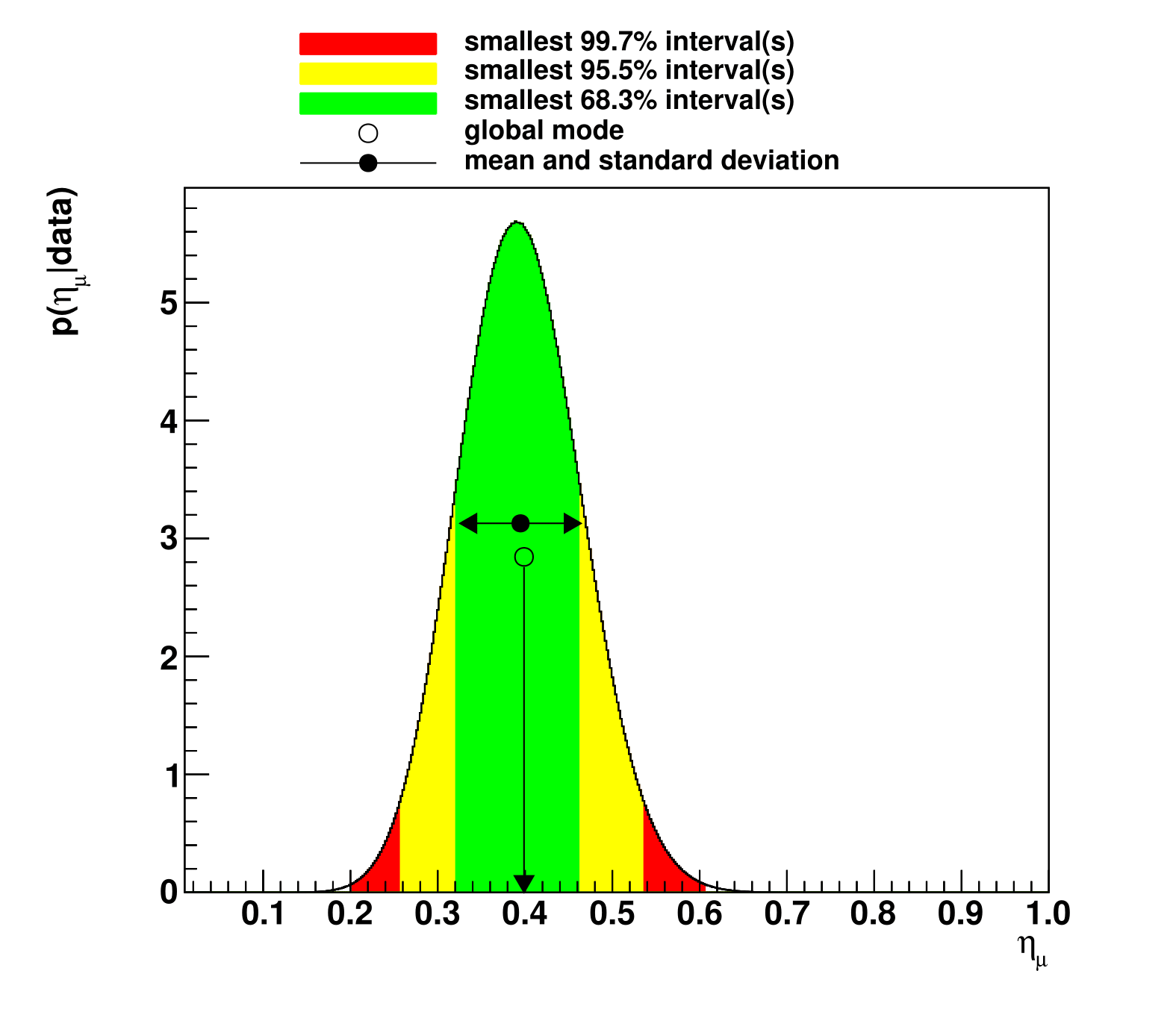} 
     \end{tabular}
     \caption{Contours of the smallest area containing 68.3\%, 95.5\%
       and 99.7\% probability of the two-dimensional marginal
       posterior distribution (left), as well as the two
       one-dimensional posterior probability densities for $\eta_{e}$
       (middle) and and $\eta_{\mu}$ (right).
     \label{fig:marginal}}
   \end{center}
 \end{figure}

 The resulting two-dimensional marginal posterior, depending on
 $\eta_{e}$ and $\eta_{\mu}$, is shown in Fig.\ref{fig:marginal},
 together with the corresponding one-dimensional projections. The
 expectation from lepton universality, i.e.,
 $\eta_{e}=\eta_{\mu}=1/3$, is indicated by the asterisk and
 consistent with the observation.

 \subsection{Inference about $k+1$ independent Poisson sources and a single observable}

 Suppose that we have an experiment in which only a single count rate
 is measured which is known to originate from $k$ different (and
 possibly correlated) background sources that are known to exist, and
 a possible additional signal source.  Taken individually, each
 background source is modeled as a Poisson process with parameter
 $b_i$, and the signal is an independent Poisson process with
 parameter $s$.

 We can write the model as we did in Section~\ref{sec-sig-k-bkg}
 above, i.e. $P(n,\vec{Z}|s,\vec{b})$, but now we have to consider
 that we do not know where the individual counts come from, i.e. we
 have to sum over all possible configurations of $\vec{Z}$. The model
 thus becomes
 \begin{eqnarray}\label{eq-single-poisson}
   P(n,\vec{Z}|s,\vec{b}) \rightarrow P(n|s,\vec{b}) & = & \sum_{\vec{Z}}  P(n,\vec{Z}|s,\vec{b}) \nonumber \\
   & = & \Poi(n | s+b)  \sum_{\vec{Z}}  \Mul(\vec{Z} \, | n, \vec{\zeta}) \\
   & = & \Poi(n | s+b) \, , \nonumber
 \end{eqnarray}
 where the latter equation follows from the normalization of the
 multinomial distribution.

 Measuring only $n$ simplifies the problem significantly, but also
 reduces the available information. The difference with respect to the
 case considered in the previous section is that the multinomial part
 has disappeared. In addition, the Poisson term now describes the sum
 of two independent sources.

 One performs inference about the signal intensity $s$ by integrating
 the marginal posterior $p(s,b|n,\alpha,\beta)$ over $b$, where
 $\alpha$ and $\beta$ describe the signal prior.  The objective result
 is the reference marginal posterior for $s$ computed in
 \citep{PoiSigBkg}
 \begin{equation}\label{eq-ref-posterior}
   p(s|n) \propto 
           e^{-s} \, f(s;n,\alpha,\beta) \, \pi(s)
 \end{equation}
 where the polynomial
 \begin{equation}\label{eq-f}
  f(s;n,\alpha,\beta) = \sum_{m=0}^{n} \binom{\alpha+m-1}{m}
                        \frac{s^{n-m}}{(n-m)! \, (1+\beta)^{m}}
 \end{equation}
 behaves like $(s+\alpha/\beta)^n$ when both $\alpha,\beta$ are very
 large.  
 Using Wolfram's \emph{Mathematica}, another representation can be found for this polynomial, in terms of the confluent hypergeometric function of the second kind $U(a,b,x)$:
 \begin{equation}
    f(s;n,\alpha,\beta) = s^n \, [(1+\beta)s]^\alpha \,
                 U(\alpha, \alpha+n+1, (1+\beta)s) / n!
 \end{equation}

 The reference prior $\pi(s)$ computed in \citep{PoiSigBkg} is an improper density, proportional to the square root of Fisher's information computed with the marginal model:
\begin{equation}\label{eq-sqrt-fisher-info}
  |I(s)|^{1/2}  =  \left|
       \left( \frac{\beta}{1+\beta} \right)^{\!\alpha} e^{-s} \,
           \sum_{n=0}^{\infty}
           \frac{[f(s;n,\alpha,\beta)]^2}{f(s;n+1,\alpha,\beta)}
       - 1
                 \right|^{1/2}
\end{equation}
 Among the methods for constructing probability matching priors considered by \citet{VCR2009}, it comes out that only the marginal model $P(n|s) = \int \Poi(n|s+b) \, \Gam(b|\alpha,\beta) \, \di b$ provides a solution: no other pseudo-likelihood can be used, as it can be checked with direct calculation.\footnote{Stefano Cabras, private communication.}

 The full objective Bayesian posterior Eqn.~\eqref{eq-ref-posterior} for
 $s$ can be computed with the help of the \emph{Bayesian Analysis
   Toolkit} \cite{BAT2009}.  However, it is often possible to adopt a quick approximation.
 Despite from its complicate form (Eqn.~\eqref{eq-sqrt-fisher-info}), it comes out that the reference prior admits a very simple limiting form that can be used in many practical problems \citep{PoiSigBkg2}.  It corresponds to the case of a perfectly known background, and performs very well in a wide portion of the $(\alpha,\beta)$ parameter space.   As a consequence, the reference marginal posterior Eqn.~\eqref{eq-ref-posterior} can
 be often well approximated by a single (truncated) Gamma density
 \citep{PoiSigBkg2}:
\begin{equation}\label{eq-lim-ref-posterior}
   p_0(s|n) = \frac{1}{C} \Gam(s+\tfrac{\alpha}{\beta}|n+\tfrac{1}{2},1)
 \end{equation}
 with normalization constant\footnote{There is a typo in the
   corresponding equation of \cite{PoiSigBkg2} which has been fixed in the corresponding erratum.}
 \begin{equation}\label{eq-lim-ref-post-norm}
   \begin{split}
   C &= \int_{0}^\infty
        \Gam(s+\tfrac{\alpha}{\beta}|n+\tfrac{1}{2},1) \, \di s
      = \int_{\alpha/\beta}^\infty \Gam(x|n+\tfrac{1}{2},1) \, \di x
  \\
     &= 1 - \frac{\gamma(n+\tfrac{1}{2},1)}{\Gamma(n+\tfrac{1}{2})}
   \end{split}
 \end{equation}
 where the last fraction defines the regularized Gamma function, that
 is the cumulative distribution function corresponding to
 $\Gam(x|n+\tfrac{1}{2},1)$, in terms of the incomplete gamma function
 $\gamma(n+\tfrac{1}{2},1)$ and the ordinary Gamma function.
 
  Incidentally, we remark the similarity between the limiting reference posterior $p_0(s|n)$ from Eqn.~\eqref{eq-lim-ref-posterior} and the reference posterior \eqref{eq-Poi-ref-post} obtained when there is no background.  Apart from the normalization, the former is the same as the latter with an offset $-\tfrac{\alpha}{\beta}$ in the origin.  This shift might be interpreted as a form of ``background subtraction'', as $\tfrac{\alpha}{\beta}$ is the prior expectation for $b$.  Thus $p_0(s|n)$ includes the particular case of no background and generalizes the result \eqref{eq-Poi-ref-post} obtained with Jeffreys' prior.

 Note also that, although the multinomial part has disappeared,
 studies of the individual background contributions are still useful:
 all background sources that have a significant impact on the total
 background contribution $b$ --- also taking into account the
 correlations on their uncertainty --- are relevant for the inference
 and the conclusions drawn from the model described in
 Eqn.~\eqref{eq-single-poisson}. One has also to be careful to
 consider any systematic effect that may have an impact on the overall
 background uncertainty.  For example, the latter could be sensitive
 to a pair of strongly anti-correlated sources of background, even
 though their variations in opposite directions have no net effect on
 $b$.  Hence they cannot be neglected in the analysis of
 systematics. In contrast those background sources, which do not
 contribute significantly to $b$ nor impact on its uncertainty, can be
 ignored in the inference about $s$.

 \subparagraph{A simple example, continued}

\begin{figure}[t]
   \begin{center}
     \includegraphics[width=0.4\textwidth]{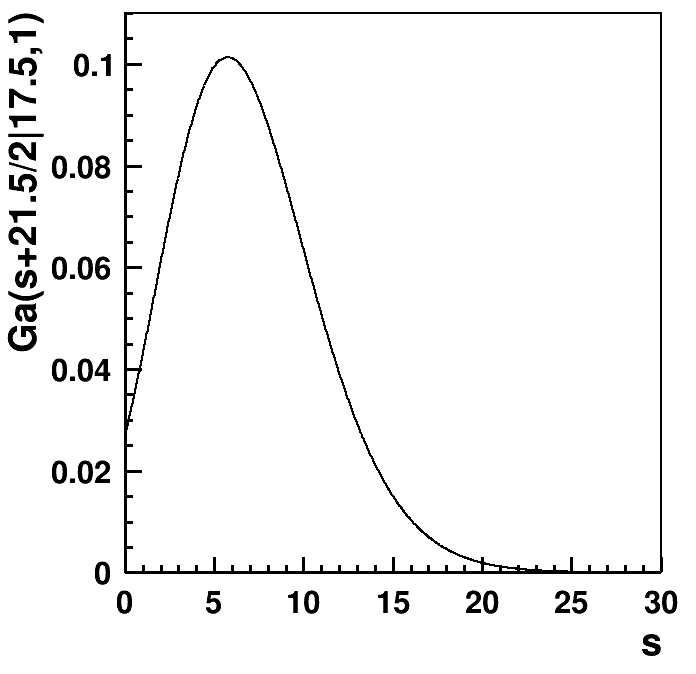}
   \caption{Marginal posterior of the signal intensity of the new radioactive source.
     \label{fig:gamma}}
   \end{center}
 \end{figure}

 After the background radioactivity in the new lab has been estimated
 by making use of the results of two experiments, a potentially
 radioactive source is brought in, to estimate its intensity.  The
 signal that we want to estimate is the intensity of the new source,
 and the measurement is performed in a laboratory in which we know
 already that some background radioactivity is detectable.  For
 simplicity, the measurement is conducted in the same time interval as
 the previous measurements.  It yields $n_{3}=17$.  Using the reference
 prior for the signal and the posterior of the second measurement of
 the background as informative prior, from
 Eqn.~\eqref{eq-lim-ref-posterior} one gets the posterior distribution
 $p_{0}(s|17) \propto \Gam(s+\tfrac{21.5}{2}|17.5,1)$ which is shown
 in Fig.~\ref{fig:gamma}.  The posterior peaks at $s=5.75$, while the
 median (mean) for the signal is at 6.6 (7.0), and the central 68.3\%
 credible interval around it (defined by the 15.85\% and 84.15\%
 posterior quantiles) is $[3.0,11.0]$.  The 95\% upper bound on the
 signal is $s < 14.2$.

\subsection{Inference about the strength of a signal distributed over several channels}

In case a signal is distributed over several channels and each of
these channels features a different background level, one can use the
likehood defined in Eqn.~\ref{eq-full-likelihood} to infer on the
signal strength. As an example, we consider the decay of a particle
into three different channels. The total number of observed 
events is $n=62$ and they are distributed among the three channels as
$n_{1}=21$, $n_{2}=29$ and $n_{3}=12$. The background contributions
are assumed to be known from auxilliary measurements and their priors
are modeled with Gamma densities with mean values and standard
deviations of $b_{1}=10 \pm 1$, $b_{2}=6 \pm 1$ and $b_{3}=2 \pm
0.5$. We use a Dirichlet prior for the branching ratios $s_{i}/s$ with
concentration parameters 0.75, 1.5 and 0.75, respectively, resulting
in expectation values of 0.25, 0.5 and 0.25 with relative
uncertainties of up to about 30\%. We choose a Jeffreys prior for the
signal contribution.

\begin{figure}[t]
   \begin{center}
     \begin{tabular}{ccc}
       \includegraphics[width=0.33\textwidth]{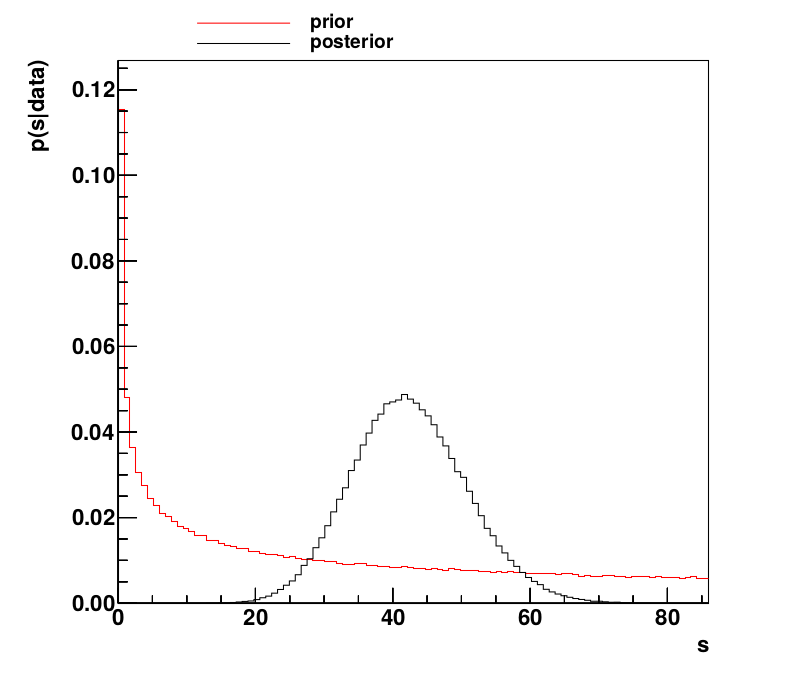} & 
       \includegraphics[width=0.33\textwidth]{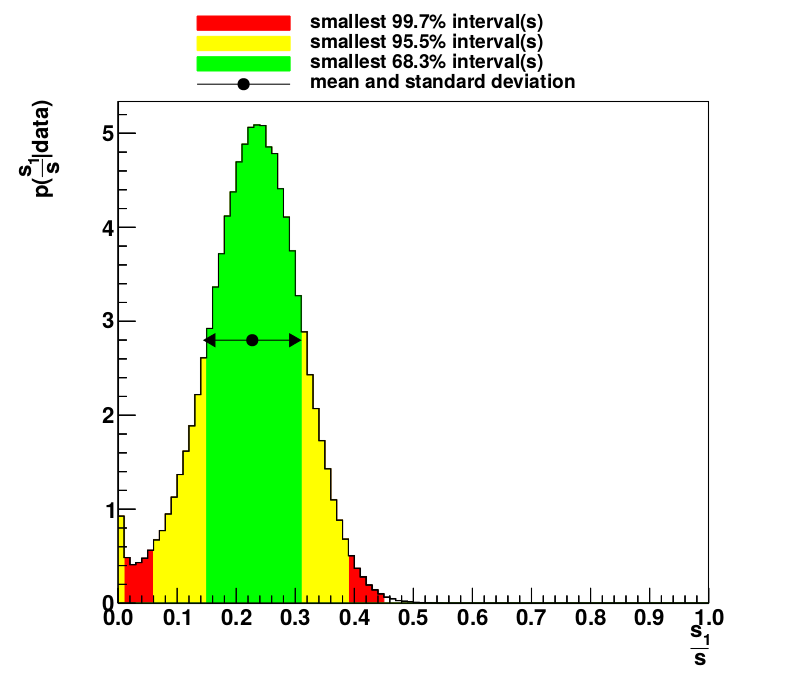} &
       \includegraphics[width=0.33\textwidth]{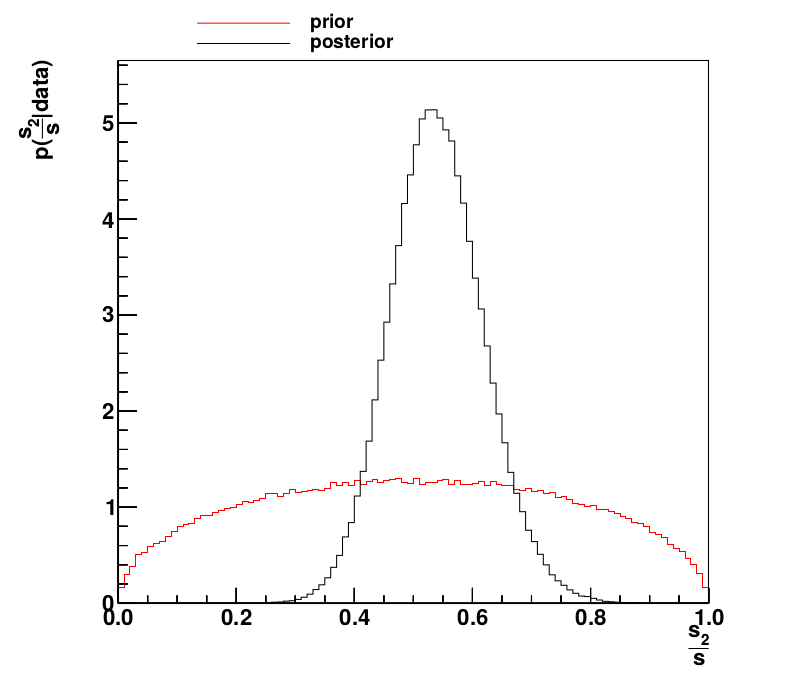} 
     \end{tabular}
     \caption{Comparison of the prior (red) and posterior probabilities (black) for the signal parameter $s$ (left) and the branching ratio $s_{2}/s$ (right). Also shown in the posterior probability density for the branching ratio $s_{1}/s$ (middle) including the smallest intervals containing 68.3\%, 95.5\% and 99.7\%.
     \label{fig:multinom}}
   \end{center}
 \end{figure}

\begin{figure}[t]
   \begin{center}
     \begin{tabular}{ccc}
       \includegraphics[width=0.33\textwidth]{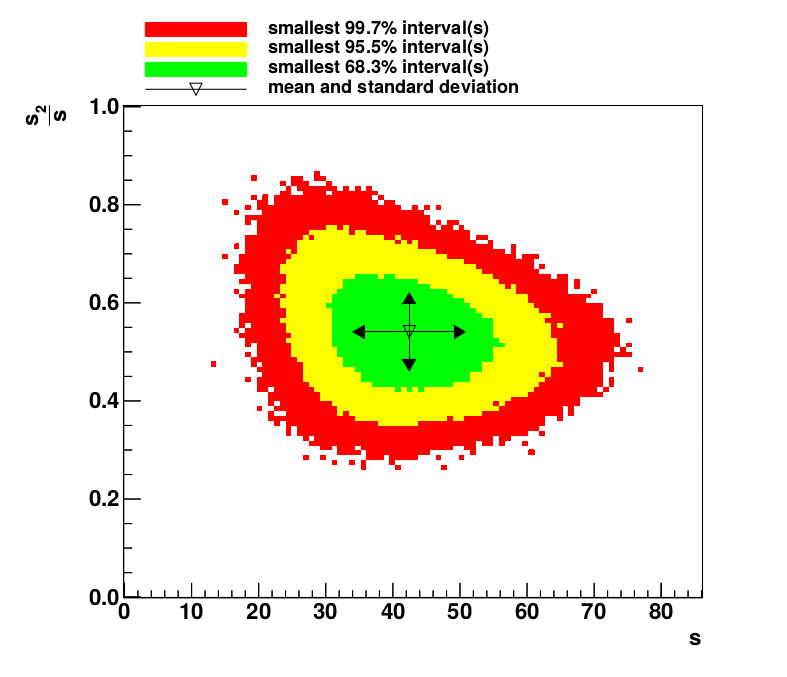} & 
       \includegraphics[width=0.33\textwidth]{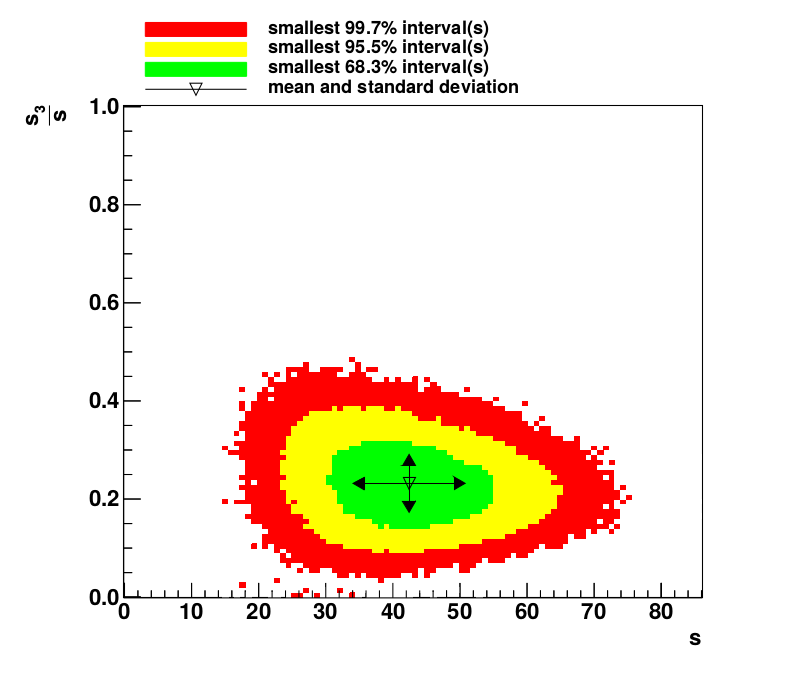} &
       \includegraphics[width=0.33\textwidth]{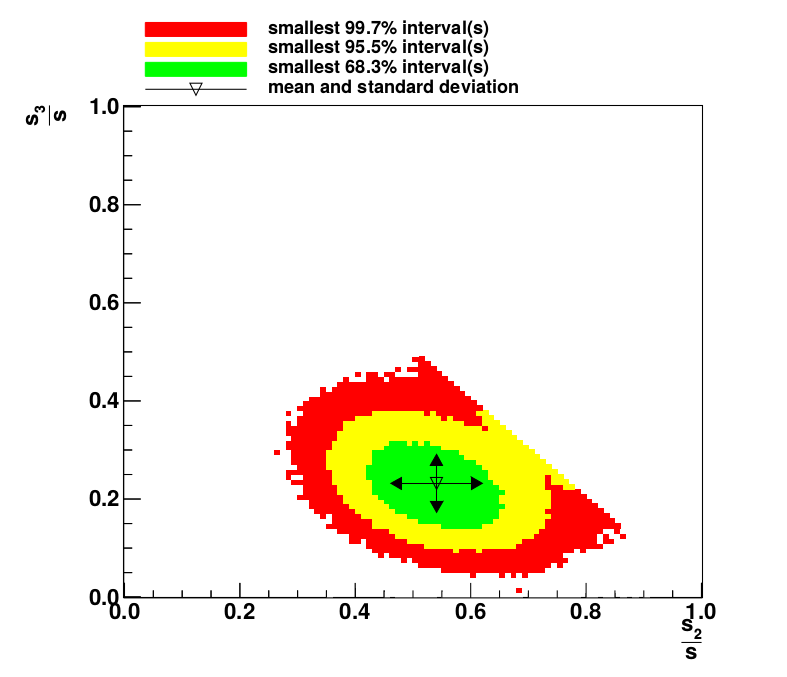} 
     \end{tabular}
     \caption{Smallest intervals containing 68.3\%, 95.5\% and 99.7\% posterior probability for the two-dimensional distributions of $s_{2}/s$ vs. $s$ (left),  $s_{3}/s$ vs. $s$ (middle), and  $s_{3}/s$ vs. $s_{2}/s$ (right). Also indicated are the mean and standard deviations.
     \label{fig:multinom_corr}}
   \end{center}
 \end{figure}

Fig.~\ref{fig:multinom} shows a comparison of the prior and posterior
probabilities for the signal parameter $s$ (left) and the branching
ratio $s_{2}/s$ (right). Also shown is the posterior probability
density for the branching ratio $s_{1}/s$ (middle). The posterior mean
and standard deviation of the marginalized posterior probability of
$s$ are 42.4 and 8.4, respectively. The mean value is close to the
mode of the distribution, and, compared to the naive expectation of $s
= n - \mathrm{E}[b]=44$, it is shifted towards smaller values due to
the steeply falling prior probability. Note that the shift is small
compared to the standard deviation. The mean values of the branching
ratio posteriors for $s_{1}/s$ and $s_{2}/s$ are 0.54 and 0.23,
respectively. The standard deviation is 0.08 in both cases, in
accordance with the expectation. As expected, the inference process
has a negligible impact on the posterior probability densities of the
background contributions. Fig.~\ref{fig:multinom_corr} shows the
smallest intervals containing 68.3\%, 95.5\% and 99.7\% posterior
probability for the two-dimensional distributions of $s_{2}/s$ vs. $s$
(left), $s_{3}/s$ vs. $s$ (middle), and $s_{3}/s$ vs. $s_{2}/s$
(right). Also indicated are the mean and standard deviations. All
three distributions show weak anti-correlations. The latter
distribution also shows a sharp cut-off at the physical boundary (the
sum of $s_{2}/s$ and $s_{3}$ can at most be unity).

 \section{Summary}\label{sec-summary}

 In this paper, we have derived a statistical model for counting
 experiments in which the sources contributing to the resulting count
 rate are potentially correlated. Using Bayes' theorem and assuming
 non-informative and informative priors on signal and background
 contributions respectively, we have shown that the marginal posterior
 probability for the signal contribution can be expressed in closed
 form. Simple use cases in the field of experimental physics have been
 presented.




\appendices

\section{Useful properties of the multinomial distribution}

 We recall here a few properties of the multinomial distribution that
 are important when considering practical problems, in particular its
 behavior under re-binning and partitioning.

 The first one is about re-binning: merging random variables
 distributed according to multinomial distributions again gives a
 multinomial distribution.  If $\vec{X} \sim \Mul(n, \vec{\eta})$ with
 $\vec{\eta} \equiv (\eta_1,\eta_2,\ldots,\eta_k)$, then $\vec{X}^*
 \equiv (X_1+X_2,X_3,\ldots,X_k) \sim \Mul(n,\vec{\eta}^*)$ where
 $\eta_1^* \equiv \eta_1+\eta_2 = (\lambda_1+\lambda_2)/\lambda$ and
 $\eta_j^* \equiv \eta_{j+1}$ for $j=2,3,\ldots,k-1$.
 One may obtain a binomial problem by sequentially merging $k-2$ bins.

 The second property is important if we know the total number of
 counts $n'$ in a subset of the bins: the conditional distribution
 given that $X_1+X_2=n'$ and $X_3+\ldots+X_k=n-n'$ is a product of two
 independent multinomial pieces:
 \begin{equation}
   \begin{split}
     (X_1,X_2) &\sim \Mul(n', \frac{\eta_1}{\eta_1+\eta_2},
                             \frac{\eta_2}{\eta_1+\eta_2})
   \\
     (X_3,\ldots,X_k) &\sim \Mul(n-n', \frac{\eta_3}{\sum_{J=3}^{k}\eta_j},
                              \ldots, \frac{\eta_k}{\sum_{J=3}^{k}\eta_j})
   \end{split}
 \end{equation}
 For example, this property tells us how to treat two histograms at
 the same time.

 Conversely, the third property is that the joint distribution of two
 independent multinomial random vectors has a multinomial kernal. To
 see this, one starts from
 \begin{equation*}
   \begin{split}
   \vec{X} &\sim \Mul(\vec{X} \, | n, \vec{\eta}) \equiv
          \frac{n!}{X_1! \, X_2! \, \cdots X_k!}
          \, \eta_1^{X_1} \, \eta_2^{X_2} \, \cdots \, \eta_k^{X_k}
   \; , \;
   \sum_{j=1}^{k} \eta_j = 1
   \; , \;
   \sum_{j=1}^{k} X_j = n
   \\
   \vec{Y} &\sim \Mul(\vec{Y} \, | n', \vec{\zeta}) \equiv
          \frac{n'!}{Y_1! \, Y_2! \, \cdots Y_m!}
          \, \zeta_1^{Y_1} \, \zeta_2^{Y_2} \, \cdots \, \zeta_m^{Y_m}
   \; , \;
   \sum_{j=1}^{m} \zeta_j = 1
   \; , \;
   \sum_{j=1}^{m} Y_j = n'
   \end{split}
 \end{equation*}
 and writes their joint distribution as
 \begin{equation}
   \label{eq-double-mul}
   \begin{split}
   \vec{X},\vec{Y} &\sim \Mul(\vec{X} \, | n, \vec{\eta}) \,
                         \Mul(\vec{Y} \, | n', \vec{\zeta})
   \\
      &= \frac{n!}{X_1! \, X_2! \, \cdots X_k!} \,
          \frac{n'!}{Y_1! \, Y_2! \, \cdots Y_m!}
         \, \eta_1^{X_1} \, \eta_2^{X_2} \, \cdots \, \eta_k^{X_k}
          \, \zeta_1^{Y_1} \, \zeta_2^{Y_2} \, \cdots \, \zeta_m^{Y_m}
   \end{split}
 \end{equation}
 Defining
 \begin{equation}
   \label{eq-multi2-vars}
   \vec{Z} \equiv (\vec{X},\vec{Y})
 \end{equation}
 such that $Z_i = X_i$ for
 $i=1,\ldots,k$ and $Z_i=Y_{i-k}$ for $i=k+1,\ldots,k+m$, one has
 \[
   \sum_{j=1}^{k+m} Z_j = n + n'
 \]
 Furthermore, defining
 \begin{equation}
   \label{eq-multi2-params}
   \vec{\xi} \equiv \frac{1}{2} (\vec{\eta},\vec{\zeta})
 \end{equation}
 one also has
 \[
   \sum_{j=1}^{k+m} \xi_j = 1
 \]
 such that Eqn.~\eqref{eq-double-mul} may be rewritten
 as
 \begin{equation}
   \label{eq-multi2}
   \begin{split}
   \vec{Z} &\sim \frac{2^{n+n'} \, n! \, n'!}{Z_1! \, Z_2! \, \cdots Z_{k+m}!} \,
         \, \xi_1^{Z_1} \, \xi_2^{Z_2} \, \cdots \, \xi_{k+m}^{Z_{k+m}}
   \end{split}
 \end{equation}
 where we used $\prod_{j=1}^{k+m} 2^{Z_j} = 2^{\sum_{j=1}^{k+m} Z_j} =
 2^{n+n'}$. 
 
  This looks like a multinomial with $k+m-1$ degrees of freedom.  However, we started from two multinomials with $k-1$ and $m-1$ degrees of freedom, hence must obtain a distribution with $k+m-2$ free parameters.  Eqn.~\eqref{eq-multi2} gives the probability distribution conditional to $n$ and $n'$.  If, however, only the total number of counts $n_\text{tot}=n+n'$ is
 known, then one must sum Eqn.~\eqref{eq-multi2} over all
 possible pairs of $n$ and $n'$ which result in a total count of $n_\text{tot}$.  This reduce by one the number of independent variables.


\end{document}